\documentclass[twocolumn]{aa}
\usepackage{graphicx}
\usepackage{natbib}
\def\gray {$\gamma$-ray\ }
\def\grays{$\gamma$-rays\ }
\def\grayse{$\gamma$-rays}

\def\Xco{$X_{\rm CO}$}

\def\Xcounitsa{molecules~cm$^{-2}$/~(K km s$^{-1}$)\ }

\def\deg{^\circ}

\def\fha{48mm}%48
\def\fhb{56mm}%61

\begin{document}
\title{The distribution of cosmic-ray sources in the Galaxy, 
\grays and the gradient in the CO-to-H$_2$ relation  }

   \subtitle{ }

   \author{A. W. Strong\inst{1},
           I. V. Moskalenko  \inst{2,3},
           O. Reimer   \inst{4},
           S. Digel   \inst{5},
           \and
           R. Diehl   \inst{1}
          }

\offprints{A. W. Strong, aws@mpe.mpg.de}

\institute{ Max-Planck-Institut f\"ur extraterrestrische Physik,
Postfach 1312, D-85741 Garching, Germany           
          %    \email{aws@mpe.mpg.de,rod@mpe.mpg.de}
\and
NASA/Goddard Space Flight Center, Code 661,  Greenbelt, MD 20771, USA 
\and
Joint Center for Astrophysics, University of Maryland, Baltimore County, 
Baltimore, MD 21250, USA
 \and
Ruhr-Universit\"at  Bochum, D-44780 Bochum, Germany
\and
W.W. Hansen Experimental Physics Laboratory, Stanford University, 
Stanford, CA 94305, USA
             }

\date{Received / Accepted  }

\abstract{
We present a solution to the apparent discrepancy between the radial
gradient in the diffuse Galactic \gray emissivity and the
distribution of  supernova remnants, believed to be the sources of
cosmic rays. Recent determinations of the pulsar distribution have
made the discrepancy even more apparent. The problem is shown to be
plausibly solved  by a variation in the $W_{\rm CO}$-to-$N$(H$_2$) scaling factor.
If this factor increases by a factor of 5--10 from the inner to the
outer Galaxy, as expected from the Galactic metallicity gradient, 
we show that the source distribution 
required to match the radial gradient of \grays can be reconciled with
the distribution of supernova remnants as traced by current studies of
pulsars.  The resulting model fits the EGRET \gray profiles
extremely well in longitude, and reproduces the mid-latitude inner
Galaxy intensities better than previous models.

   }
\titlerunning{Distribution of cosmic ray sources in the Galaxy}
\authorrunning{Strong A.W. et al.}
\maketitle
%
%________________________________________________________________

\section{Introduction}

The puzzle of the Galactic \gray gradient goes back to the time of the COS-B
satellite \citep{bloemen86,strong88}; using HI and CO surveys to trace the atomic and
molecular gas,  the Galactic distribution of emissivity per H atom is a
measure of the cosmic-ray (CR) flux, for the gas-related bremsstrahlung and
pion-decay components. However the gradient  determined in this way is much
smaller than expected if supernova remnants (SNR) are the sources of cosmic
rays, as is generally believed. This discrepancy was confirmed with the
much more precise data from EGRET on the COMPTON Gamma Ray Observatory,
even allowing for the fact that inverse-Compton emission (unrelated to
the gas) is more important than originally supposed \citep{SMR00}.
A possible explanation of the small gradient in terms of CR propagation, 
involving radial variations of a Galactic wind,  was recently put forward by 
\citet{breitschwerdt02}.

\begin{figure}
%   \centering     \includegraphics[width=65mm,height=60mm]{fig1.ps}
   \centering     \includegraphics[width=0.99\linewidth,height=\fha]{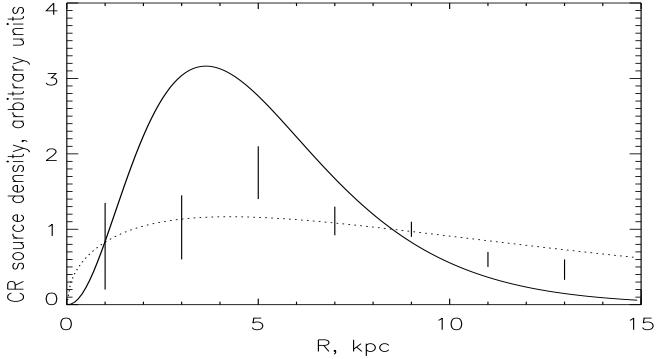}
   \caption{CR source density as function of Galactocentric
radius $R$.  Dotted: as used in \citet{SMR00}, solid line:based on
pulsars \citep{lorimer04}  as used in this work, vertical bars: SNR
data points from \citet{case98}.  Distributions are normalized at $R =
8.5$ kpc.}  \label{fig1}
   \end{figure}

However the derivation of the Galactic distribution of SNR, 
commonly based on radio surveys, is subject to large
observational selection effects, so that it can be argued that the
discrepancy is not so serious. But other tracers of the distribution
of SNR are available, in particular pulsars; the new sensitive Parkes
Multibeam survey with 914 pulsars has been used by \citet{lorimer04}
to derive the Galactic distribution, and this confirms the
concentration to the inner Galaxy.  Fig.~\ref{fig1} compares the
pulsar distribution from \citet{lorimer04} with a CR source
distribution  which fits the EGRET \gray data \citep{SMR00}.  If the
pulsar distribution indeed traces the SNR, then  there is a serious
discrepancy with \grayse.  The distribution of SNR given by
\citet{case98} is not so peaked, but the number of known SNR is much
less than the number of pulsars and the systematic effects very
difficult to account for \citep{green96}. But even this flatter
distribution is hard to reconcile with that required for \grayse.
Another, quite independent, tracer of the SNR  distribution is the
1809 keV line of $^{26}$Al; whether this originates mainly in type II
supernovae or masssive stars is not important in this context, since
both trace star-formation/SNR. The COMPTEL  $^{26}$Al maps
\citep{knodlseder99,pluschke01} show that the emission is very
concentrated to the inner radian of the Galaxy.  The density of free
electrons shows a similar distribution \citep{cordeslazio03}.  The
$^{26}$Al measurements are not subject to the selection effects of
other methods; although they have their own uncertainties, they
support  the type of distribution which we adopt in this paper.

%-----------------------------------------------------------
%______________________________________________________________

A major uncertainty in the models of diffuse Galactic \gray emission
is the  distribution of molecular hydrogen, as traced by the
integrated intensity of the $J$ = 1--0  transition of $^{12}$CO,
$W_{\rm CO}$.  Gamma-ray analyses have in fact provided one of the
standard values for the scaling factor\footnote{units: \Xcounitsa}
\Xco = $N(H_2)/W_{\rm CO}$; with only the assumption that cosmic rays
penetrate molecular clouds freely, the \gray values are free of the
uncertainties of other methods (e.g. those based on the assumption of
molecular cloud
 virialization). However previous analyses, e.g.
\citet{strongmattox96}, \citet{hunter97}, \citet{SMR00}, have usually assumed that
\Xco\ is independent of Galactocentric radius $R$, since otherwise the
model has too many free parameters. But  there is now good reason to
believe that \Xco\ increases with $R$, both from COBE/DIRBE studies
\citep{sodroski95,sodroski97} and from the measurement of a Galactic
metallicity gradient combined with the strong inverse dependence of
\Xco\ on metallicity in external galaxies  \citep{israel97,israel00}.
A rather rapid radial variation of \Xco\ is expected,  based on a
gradient in [O/H] of 0.04--0.07 dex/kpc
\citep{hou00,deharveng00,rolleston00,smartt01,andrievsky02} and the
dependence of \Xco\ on metallicity in external galaxies: $\log$
\Xco$\propto -2.5$   [O/H] \citep{israel97,israel00}, giving \Xco
$\propto 10^{(-0.14\pm0.04)R}$, amounting to a factor 1.3--1.5 per
kpc, or an order of magnitude between the inner and outer Galaxy.
\footnote{The values given by \citet{israel97,israel00} include the
effects of the radiation field, implicitly containing the radiation
field/metallicity correlation of his galaxy sample. \Xco\ is
positively and almost linearly correlated with radiation field, so the
dependence of \Xco\ for constant radiation field is even larger: $\log$
\Xco$\propto -4$  [O/H] \citep{israel00}. By adopting the coefficient
--2.5 we implicitly assume the same radiation/metallicity correlation
within the Galaxy as over his galaxy sample.}  A less rapid
dependence, $\log$ \Xco$\propto$--1.0  [O/H], was found by
\citet{boselli02}, which however still implies a significant \Xco($R$)
variation.  \citet{boissier03} also combine the metallicity gradient
with \Xco$(Z)$ within individual galaxies, to obtain radial profiles
of H$_2$, and give arguments for the validity of this procedure.
\citet{digel90} found that molecular clouds in the outer Galaxy 
($R$$\sim$12 kpc) are underluminous in CO, with \Xco\ a factor 4$\pm$2
times the  inner Galaxy value.  \citet{sodroski95,sodroski97} derived a
similar variation ($\log$\Xco/10$^{20} = 0.12R - 0.34$) when  modelling
dust emission for COBE data. \citet{pak98} predicted the physical origin  for a variation of \Xco\ with Z. \citet{papadopoulos02}
and \citet{papadopoulos04} discuss  the physical state of this metal-poor gas phase
in the outer parts of spiral galaxies (relatively warm and diffuse).  Observations of H$_2$ line emission from NGC~891
with ISO \citep{valentijn99}  indicate a massive cool molecular
component in the outer regions of this galaxy, supporting the trend
found in our Galaxy.
 
%----------------------------------------------------------- 
   \begin{figure}
   \centering    
   \includegraphics[width=0.99\linewidth,height=\fha]{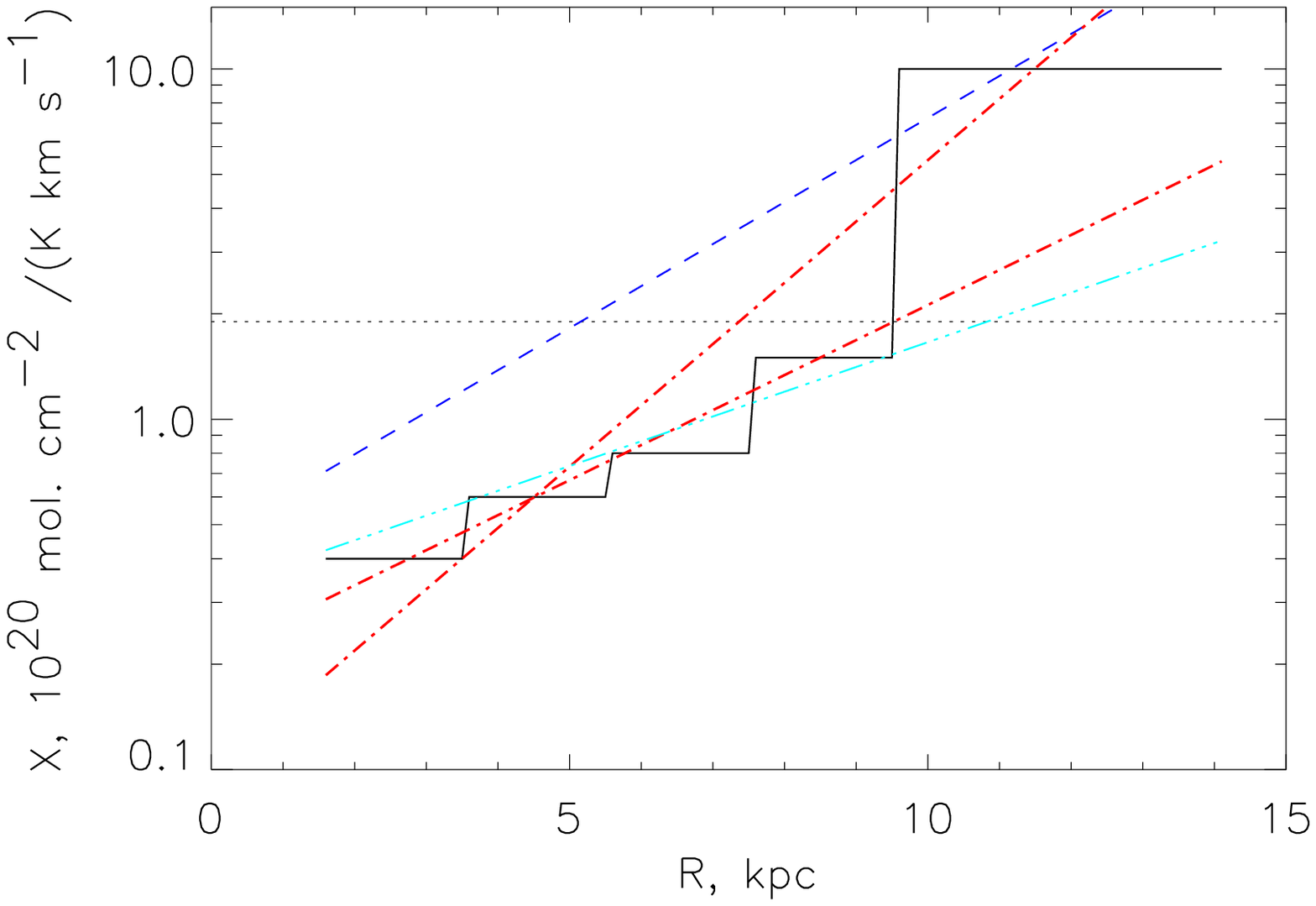}
   \caption{\Xco\ as function of  $R$.
    Dotted horizontal line, black: as used in \citet{strongmattox96,SMR00}; 
    solid  line, black: as used for \grays in this work; 
    dashed, dark blue: from \citet{sodroski95};
    dash-dot, red: using metallicity gradient as described in the text, 
    \Xco$\propto Z^{-2.5}$ \citep{israel00}, two lines
    for [O/H] = 0.04 and 0.07 dex/kpc;
    dash-dot-dot,light blue: using \Xco$\propto Z^{-1.0}$ \citep{boselli02}
    and [O/H] =  0.07 dex/kpc.
    The values using metallicity are normalized approximately to those 
    from the \gray analysis.} \label{fig2}
   \end{figure}
%______________________________________________________________

Fig.~\ref{fig2} illustrates some of the possible \Xco\ variations
implied by these studies.  For the cases where metallicity is used to
estimate  \Xco, the values are normalized approximately to the
values used in the present \gray analysis, since we are only
interested in comparing the variations of \Xco.  From the viewpoint of
\grayse, the effect of a steeper CR source distribution is
compensated by the increase of \Xco. Thus we might expect to resolve
the  apparent discrepancy in the source distribution, and improve our
understanding of the Galactic \gray emission. In this paper we
investigate quantitatively this possibility.  Note that the \grays
include major  contributions from interactions with atomic hydrogen
and from inverse Compton scattering, both of which are independent of
\Xco; this means that the \Xco\ variation has to be quite large to
have a significant effect.

%-----------------------------------------------------------------
\section{Data}

The EGRET and COMPTEL data are the same as described in
\citet{SMR00,SMR04a}.  The EGRET data consist of the standard product
counts and exposure for 30 MeV -- 10 GeV, augmented with data for 10--120
GeV. The \gray point sources in the 3EG catalogue have been
removed as described in \citet{SMR00}.  The HI and CO data are as
described in \citet{moskalenko02} and \citet{SMR04a}; they consist of combined
surveys divided into 8 Galactocentric rings on the basis of kinematic
information.  Full details of the procedures for comparing models with
data are given in \citet{SMR04a} to which the reader is referred.

%----------------------------------------------------------- 
   \begin{figure}
   \centering    
   \includegraphics[width=0.99\linewidth,height=\fhb]{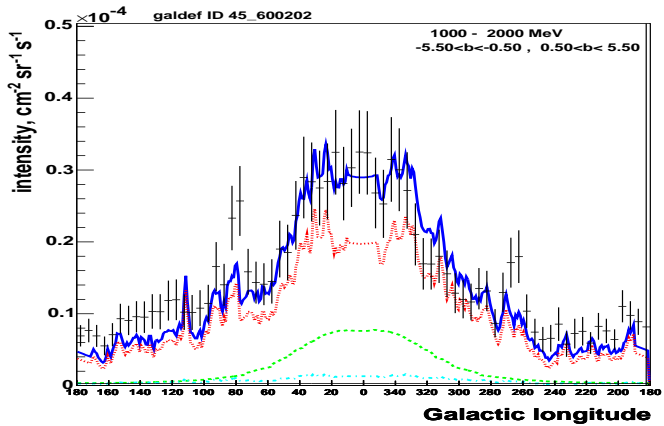}
   \caption{Longitude profile of \grays for 1000--2000 MeV, averaged over 
$|b|<5.5\deg$. 
Vertical bars: EGRET data; lines are model components convolved with the EGRET 
point-spread function:
green: inverse Compton emission, red: $\pi^0$-decay, light blue: bremsstrahlung,
dark blue: total.} \label{fig3}
   \end{figure}
%______________________________________________________________

%----------------------------------------------------------------------
\section{Model and Method}
We use the GALPROP program \citep{SMR00,SMR04a} to compute the models.
GALPROP was extended to allow a variable \Xco($R$) to be input.  The
distribution of CR sources is assumed to follow that of
pulsars in the form given by \citet{lorimer04}, as shown in
Fig.~\ref{fig1}. The other parameters, in particular the CR
nucleon and electron injection spectral shape and propagation
parameters, are taken from the ``optimized model'' of
\citet{SMR04a}.  As before the halo height is taken as $z_h$ = 4 kpc,
and the maximum radius $R$ = 20 kpc.  The isotropic background is as
derived in \citet{SMR04b}.  Since in this work we simply wish to
demonstrate the possibility to obtain a plausible solution, we adopt a
heuristic approach, adjusting \Xco($R$)  to obtain a satisfactory
solution as shown in Fig.~\ref{fig2}.  The electron flux has been
scaled down by a factor 0.7 relative to \citet{SMR04a} to obtain an
optimal fit.

%-----------------------------------------------------------------------------
\section{Results}
\begin{figure}
   \centering    
   \includegraphics[width=0.99\linewidth,height=\fhb]{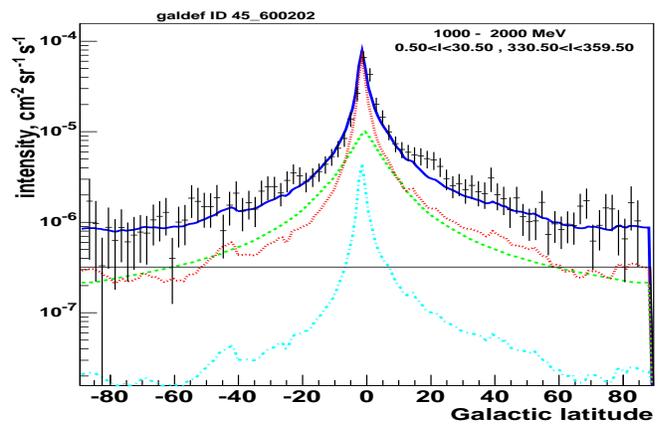}
   \caption{Latitude profile of \grays for 1000--2000 MeV, averaged over 
$330\deg<l<30\deg$. Data and curves as in Fig.~\ref{fig3}.  
The extragalactic background is shown  as a black horizontal line. } \label{fig4}
   \end{figure}

\begin{figure}
   \centering    
   \includegraphics[width=0.99\linewidth,height=\fhb]{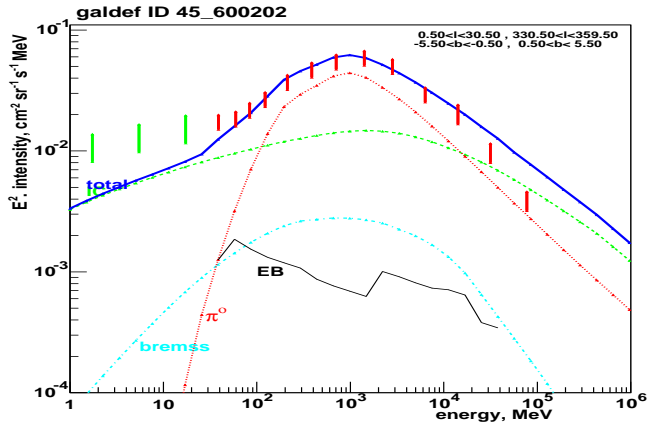}
   \caption{Spectrum of inner Galaxy, $330\deg<l<30\deg, |b|<5.5\deg$.
Vertical bars: EGRET data  (red), COMPTEL data (green). 
Curves: predicted intensities; inverse Compton (green), 
$\pi^0$-decay (red), bremsstrahlung (light blue), extragalactic background 
(black), total (dark blue). } \label{fig5}
   \end{figure}

Figs.~\ref{fig3} and \ref{fig4} show the longitude and latitude
distributions  for 1--2 GeV, compared to EGRET data.  A rather rapid
variation of \Xco\ is required to compensate the CR source gradient,
but it is fully compatible with the expected variation based on
metallicity gradients and the COBE result, as described in the
Introduction.  The longitude and latitude fits are good except in the
outer Galaxy where the prediction is rather low. One possible reason
for this is that the CR source density does not fall off so fast
beyond the Solar circle as given by the adopted pulsar distribution,
which has an exponential decay.  Another possibility could be even
larger amounts of H$_2$ in the outer Galaxy than we have assumed (see
discussion in Introduction).  We have chosen the range 1--2 GeV for
the profiles since this is where the gas contribution and hence the
effect of \Xco\ is maximal.  An exhaustive comparison of profiles in
all energy ranges is beyond the scope of this Letter, but in fact the
agreement is good at all energies.  The larger CR gradient in this
model has another consequence:  the predicted inverse-Compton emission
in the inner Galaxy  is more intense  at intermediate latitudes where
the interstellar radiation field is still high; this is precisely the
region where previous models \citep{hunter97,SMR00,SMR04a} have had
problems to reproduce the EGRET data.  Fig.~\ref{fig5} shows the model
spectrum of the inner Galaxy compared with EGRET data; the fit is
similar to that of  models \citep{SMR04a} with {\it ad hoc} source
gradient and constant \Xco. The prediction is rather high above 20
GeV, however the EGRET data are least certain in this range
\citep{SMR04a}.
  
%----------------------------------------------------------- 
%______________________________________________________________

%------------------------------------------------------------------------------
\section{Discussion}
We have shown that a good fit to the EGRET data is obtained with the
particular  combination of parameters chosen.  We can however ask
whether the pulsar source distribution combined with a constant  \Xco\
could also give a good fit if we reduce the CR electron intensity, to
supress  the inner Galaxy peak from inverse Compton emission. This can
indeed reproduce the  longitude profile in the inner Galaxy, but fails
badly to account for the latitude  distribution, since it has a large
deficit at intermediate latitudes.  Some variation of \Xco\ is
therefore required.  The suggested variation of \Xco\ would have
significant impact on the  Galactic H$_2$  mass and distribution.
Warm molecular hydrogen in the outer parts of spiral  galaxies that is
not traced by CO emission may be detectable by the Spitzer
observatory in 28 $\mu$m vibrational emission.  These issues will be
addressed in future work.
 
%----------------------------------------------------------- 
%______________________________________________________________

%------------------------------------------------------------------------------
\section{Conclusions}
Two {\it a priori} motivated developments allow us to obtain a more
physically plausible model for Galactic \grayse, simultaneously
allowing a CR source distribution similar to SNR as traced by pulsars
and an expected variation in the $W_{\rm CO}$-to-$N$(H$_2$) conversion
factor.  Obviously the uncertainty in both the source distribution and
\Xco\ are large so our solution is far from unique, but it
demonstrates the possibility to obtain a physically-motivated model
without resorting to an {\it ad hoc} source distribution. This result
supports the SNR origin of CR.  The resulting model also gives
improved predictions for \grays in the inner Galaxy at mid-latitudes.
We have therefore  achieved a step towards a better understanding of
the diffuse Galactic \gray emission.  This result is important input
to the development of models for the upcoming GLAST mission.   This
Letter is intended only to point out the potential importance of the
effect.  The next step will be a more quantitative analysis to derive
\Xco($R$) from the \gray data themselves.

%______________________________________________________________
\begin{acknowledgements}
We  thank F.~Israel and D.~Lorimer and the referee for useful discussions.
I.V.M.\ acknowledges partial support from a NASA Astrophysics Theory Program
grant, O.R.\ acknowledges 
support from the BMBF through DLR grant QV0002.
 
%of his \Xco - metallicity relation to the Galaxy.
% and D. Lorimer for
%discussion of the pulsar distribution. 
%We thank the referee for pointing out additional relevant  material
%and  implications of this work, in particular the relevance for the 
%Spitzer observatory.

\end{acknowledgements}

%%%%%%%

\end{document}